# Periodic Ridged Leaky-Wave Antenna Design Based on SIW Technology

Sajad Mohammad-Ali-Nezhad and Alireza Mallahzadeh, *Member, IEEE*

*Abstract*—A periodic long slot leaky-wave antenna (LWA) based on a ridged substrate integrated waveguide (RSIW) is proposed. To reduce the cross polarization, the long slot is placed on the SIW centerline, and a sinusoidal ridge is used to produce a controllable asymmetric electric field around the long slot. Also, only a small stopband occurs when the beam is scanned through the broadside. Varying the amplitude and width of the sinusoidal ridge provides a fixed phase constant and controllable leakage rate to achieve the desired sidelobe level of less than $-30$ dB. Measurement results are consistent with the simulation results.

## I. INTRODUCTION

LONG slot leaky-wave antennas (LWAs) are simple structures with many applications that work as uniform leaky-wave antennas [1]–[4].

The desired far-field pattern of the straight long slot LWA can be achieved by controlling the aperture field distribution of the long slot. Although properly tapering the slot position improves the sidelobe level (SLL), this method can produce second-order beams and increase the cross polarization [4].

This problem can be solved by using a ridge in slotted waveguide antennas. Asymmetric ridged waveguide, convex and concave double-ridged waveguide, and tilted single-ridged waveguide with centered slots are applied for standing wave antennas [5]–[7]. Also, cross polarization can be reduced using parallel-plate waveguide (PPW) stubs [8]–[10]. However, the presence of these stubs in the structures increases their height.

In [11], a wiggly ridged long slot waveguide LWA is proposed for realizing a low cross-polarization traveling-wave antenna. In leaky-wave antennas, the SLL can be controlled by tapering the leakage rate along the antenna length; also, all parts of the antenna aperture should radiate at the same angle that requires an unchanged phase constant along all the antenna lengths [1]. In the structure proposed in [11], changing the ridge position can vary both leakage rate and phase constant simultaneously, resulting in a wide radiation pattern. Another disadvantage of this antenna is in scanning applications in which the stopband occurs when the beam is scanned through the broadside.

Among LWAs, periodic leaky-wave antenna (PLWA) structures have attracted a lot of attention due to their beam-scanning ability from backward to forward quadruples. In [12], an LWA based on a periodic set of slots printed on a dielectric waveguide that are modulated to achieve low-SLL scanning is presented.

In recent years, various SIW leaky-wave antennas have been proposed, one of which is the SIW long slot leaky-wave antenna [13]–[16]. In [16], the straight slot etched on the broadside of a meandering SIW shows that the desired SLL can be obtained by tapering the aperture distribution of the centered slot. As the slot is designed to be symmetrical, the SIW straight long slot LWA can improve the cross-polarization level up to $-30$ dB. In such structures, changing the value of slot distance from the sidewall can vary both leakage rate and phase constant simultaneously. However, designing these structures based on PPW stubs could result in structures with a fixed phase constant and variable leakage rate. Consequently, beam broadening in the antenna patterns could be prevented [8]–[10].

In this letter, a periodic LWA based on a sinusoidal ridged SIW (RSIW) has been proposed. In the suggested structure, the stopband in the broadside is somehow diminished. Simultaneously varying the amplitude and width of the sinusoidal ridge results in a variable leakage rate and unchanged phase constant to obtain the desired radiation pattern. Since the long slot is located in the center of the top plane of the SIW, the antenna structure is symmetrical, and the cross polarization is proper as well.

## II. PRINCIPLE OF OPERATION OF SINUSOIDAL RSIW-BASED LWA

When the slot is distanced from the center of the upper plane of the SIW, cross polarization increases due to the asymmetric electric field distribution on both sides of the slot, but radiation becomes possible. In order to improve the cross polarization, the slot has to be located in the center, but the electric field distribution should be varied to become asymmetric around the slot.

Using the ridge in the SIW causes the electric field to be concentrated between the ridge and the upper plane of the SIW. In this situation, for keeping the cross polarization at an acceptable level, the slot is located in the center, and also for making radiation possible from the slot, the ridge shape and position are varied.

Another advantage of the ridge is that it increases the bandwidth, where in LWA structures, this wide bandwidth can help control antenna features.

In [17], the sinusoidal microstrip LWA has been used for the microwave field focusing application. Sinusoidal ridges can also be used inside SIW structures for controlling antenna properties, as shown in Fig. 1.

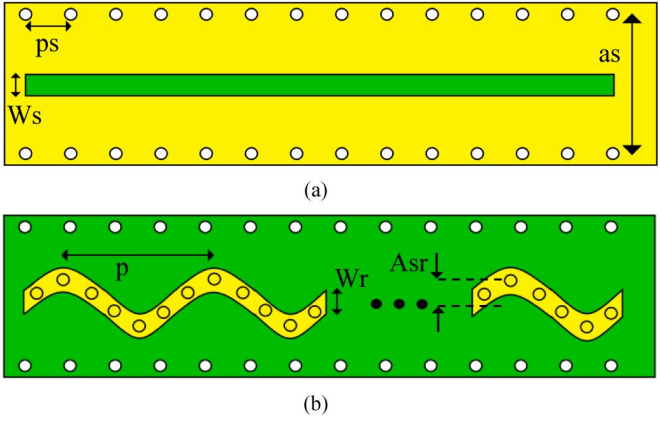

Fig. 1. Geometry of the sinusoidal ridged SIW long slot LWA. (a) Top layer. (b) Middle layer.

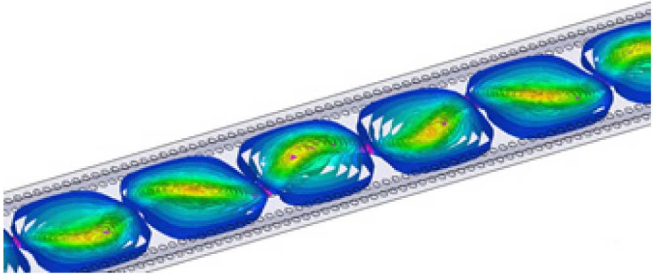

Fig. 2. Electric field distribution of the proposed structure in the broadside.

Fig. 2 shows that the electric field is located between the ridge and the broad wall of the SIW and varies as a response to the variations of the ridge in the SIW, hence it is asymmetric to the center of the SIW's upper plane and makes radiation possible for the centered slot in the broad wall of the SIW.

The periodic LWA suffers a stopband in the broadside. However, in the proposed antenna, because of smooth and sinusoidal variations of the ridge, the Bloch impedance can be better controlled, thus a better result is obtained in comparison to [8].

Leakage rate ($\alpha$) and phase constant ($\beta$) can be controlled by changing the amplitude ($Asr$), width ($Wr$), height ($h$), and period of the sinusoidal ridge ($p$). In order to avoid the increase in the layers of the SIW and fabricating the structure with maximum two substrates, the height can be kept fixed, so for achieving the desired radiation properties, other ridge features can be varied. $\beta$ can be controlled based on the variations of the ridge period. A change in the amplitude of the sinusoidal ridge can vary $\alpha$; it also changes $\beta$ slightly. For keeping $\beta$ unchanged, the width of the ridge has to be controlled.

## III. PROPAGATION BEHAVIOR OF THE PERIODIC LWA BASED ON THE SINUSOIDAL RSIW

The inclusion of the periodic sinusoidal ridge along the RSIW excites an infinite number of space harmonics in the structure, each characterized by phase constants $\beta_n$

$$\beta_n = \beta_0 + \frac{(2n+1)\pi}{p} \quad (1)$$

where $p$ is the length of the period of the sinusoidal ridge in Fig. 1 and $\beta_0$ is the phase constant of the dominant mode of the RSIW.

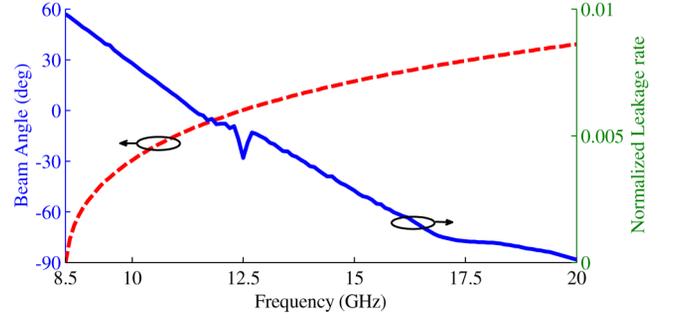

Fig. 3. Leakage rate and beam angle variations versus frequency ($b1 = 30$ mil).

The pointing angle of the radiated beam from the broadside direction and the 3-dB beamwidth for an LWA with a length of $L_A$ can be determined as

$$\sin \theta_m \cong \frac{\beta_{-1}}{k_0} \quad (2)$$

$$\Delta\theta \cong \frac{1}{\frac{L_A}{\lambda_0} \cos \theta_m} \approx \frac{\alpha/k_0}{0.183 \cdot \cos \theta_m} \quad (3)$$

where $\lambda_0$ is the free-space wavelength, and the length of the antenna has been chosen so that 90% of the power is radiated.

The complex propagation behavior of the proposed LWA is obtained using HFSS software based on the method presented in [14]. Geometrical variations of the antenna in different frequencies result in suitable leakage rates and phase constants that can be used to control the structure to achieve the desired radiation pattern.

### A. Variations of $\alpha$ and $\beta$ With Frequency

Through analyzing the changes in $\alpha/k_0$ and $\beta/k_0$, caused by frequency variations, the possible radiation frequency range can be found. In other words, the dimensions for the structure, which make the desired radiation pattern possible within the operating frequency range, are selected.

Variations of the beam angle and $\alpha/k0$ based on the variation of frequency are shown in Fig. 3. The antenna main beam scans with frequency changes; in fact, it moves from backward endfire to the forward quadrant as the frequency increases from 8.5 to 20 GHz. Around the broadside, the beam angle is linear and $\alpha/k0$ is relatively constant.

### B. Dependence of $\alpha$ and $\beta$ on Geometrical Parameters

The variation of antenna dimensions would vary $\alpha/k0$ and $\beta/k0$, thus proper geometrical dimensions should be found for the desired radiation pattern. Fig. 4 shows the variations of $\beta/k0$ and $a/k0$ when the amplitude of the sinusoidal ridge ($Asr$) varies from 0 to $b$ at 10 GHz. The leakage rate can be controlled to have values with in the range of zero to a maximum value. $\alpha$ reaches its maximum value when $Asr$ has its maximum value, i.e., when the sinusoidal ridge has its maximum distance from the center of the SIW. Also, when $Asr = 0$, the slot does not radiate, which is equal to the situation in which the ridge is located in the center and no radiation takes place due to the symmetry in the electric field distribution.

The behavior of leakage and phase constants as a function of the ridge width is shown in Fig. 5. The value of $\alpha/k0$ varies

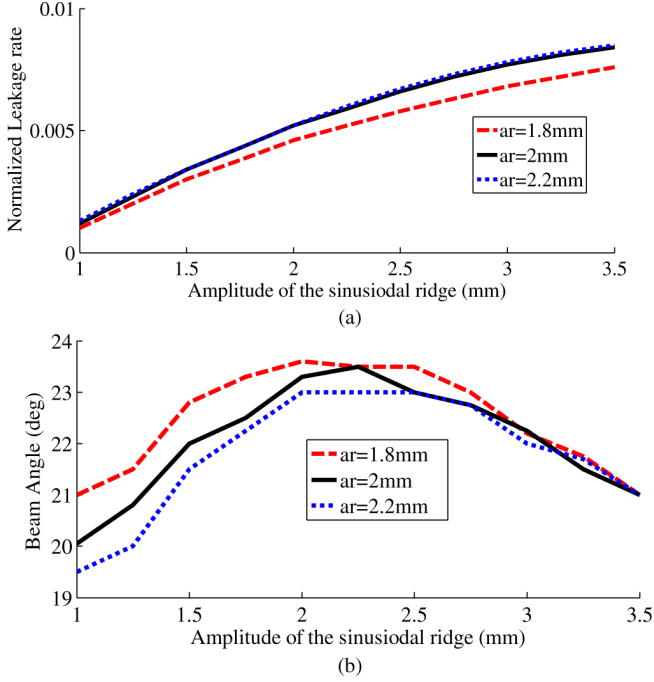

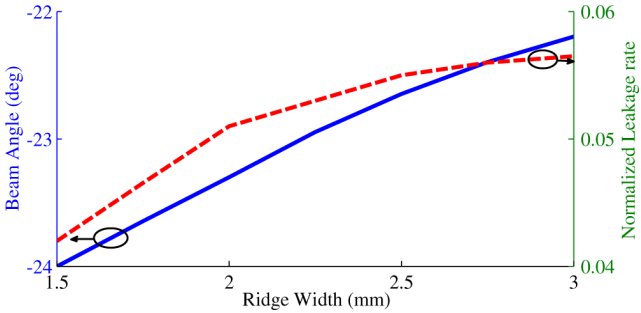

Fig. 4. (a) Normalized leakage constant and (b) beam angle for the LWA based on sinusoidal RSIW versus the amplitude of the sinusoidal ridge for several values of ridge widths at 10 GHz.

Fig. 5. Normalized leakage constant and beam angle for the LWA based on sinusoidal RSIW versus the ridge width of the sinusoidal ridge at 10 GHz.

smoothly for different values of $Wr$. Variations of $Wr$ can also vary $\beta/k0$.

## IV. TAPERED ANTENNA DESIGN FOR SLL REDUCTION

The aperture distribution of the long slot is set to follow the Taylor distribution with SLL $= -30$ dB. There is a mathematical relationship between the leakage rate along the antenna length $\alpha(z)$ and the desired amplitude distribution $A(l)$ [1]

$$\alpha(z) = \frac{\frac{1}{2}|A(l)|^2}{\frac{1}{1-R}\int_0^L |A(z)|^2 dz - \int_0^l |A(z)|^2 dz}. \quad (4)$$

Varying $Asr$ and $Wr$ would vary both $\alpha$ and $\beta$ simultaneously, while to obtain the desired pattern, these parameters should change the value of $\alpha$ but maintain an unchanged $\beta$.

In order to solve this problem, we can vary both $Asr$ and $Wr$ simultaneously at any point of the aperture, where the radiation is stabilized in a fixed direction and the leakage constant would be desirable to realize a proper SLL.

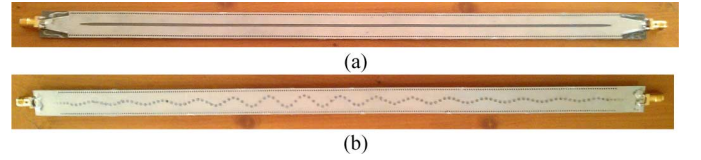

Fig. 6. Photographs of the proposed LWA based on the sinusoidal RSIW. (a) Front view. (b) Back view.

TABLE I
VALUES OF $Asr$ AND $ar$ FOR SLL $= -30$ dB AND $\theta_m = -30°$

| position | Asr(mm) | ar(mm) | position | Asr(mm) | ar(mm) |
|---|---|---|---|---|---|
| 0.5p | 0.75 | 1.6 | 1p | 0.77 | 1.62 |
| 1.5p | 0.8 | 1.65 | 2p | 0.84 | 1.7 |
| 2.5p | 0.9 | 1.8 | 3p | 1 | 1.88 |
| 3.5p | 1.12 | 1.95 | 4p | 1.25 | 2.35 |
| 4.5p | 1.4 | 2.4 | 5p | 1.57 | 2.6 |
| 5.5p | 1.75 | 2.8 | 6p | 1.95 | 2.8 |
| 6.5p | 2.2 | 2.8 | 7p | 2.45 | 2.75 |
| 7.5p | 2.7 | 2.5 | 8p | 2.98 | 2.4 |
| 8.5p | 3.25 | 2.4 | 9p | 3.23 | 2.65 |
| 9.5p | 3 | 2.65 | 10p | 2.7 | 2.8 |
| 10.5p | 2.3 | 2.8 | 11p | 2 | 2.3 |
| 11.5p | 1.6 | 2 | 12p | 1.3 | 1.8 |
| 12.5p | 1.1 | 1.8 | 13p | 1 | 1.75 |
| 13.5p | 0.95 | 1.73 | 14p | 0.91 | 1.71 |
| 14.5p | 0.89 | 1.7 | 15p | 0.87 | 1.69 |

In [9], the method for producing a graph that is used to determine the dimensions of the structure for realizing different values of $\alpha$ and a fixed $\beta$ is presented. Table I is created for the proposed structure based on this graph. In Table I, values of $Asr$ and $Wr$ at each part of the sinusoidal ridge of the LWA are provided to obtain the desired radiation pattern with an SLL of $-30$ dB, based on (4) with $\theta_m = -30°$.

## V. MEASUREMENT RESULTS

The two-layer LWA based on the RSIW was fabricated using a Rogers 5880 substrate with a thickness of $h = 0.8$ mm and a relative permittivity of $\varepsilon_r = 2.2$. A photograph of the proposed antenna is shown in Fig. 6. A tapered microstrip line has been used to feed the proposed antenna. For impedance matching, the microstrip line has been tapered to match a 50-$\Omega$ coaxial cable to the input impedance of the RSIW. When the tapered microstrip line reaches its middle, the presence of the ridge in the middle layer starts with a width that is smoothly tapered from 0 to $Wr$ in order to make field matching possible.

The measured and simulated co- and cross polarizations in the normalized radiation pattern of the proposed LWA at 9, 10, and 11 GHz are shown in Fig. 7; also, beam steering from $-30°$ to $-20°$ can be observed. The measured SLL and cross polarization are lower than 30 and 50 dB, respectively, which coincide with the simulated results.

The measured input reflection coefficient of the proposed LWA is shown in Fig. 8. In the 8.5–16-GHz frequency band,

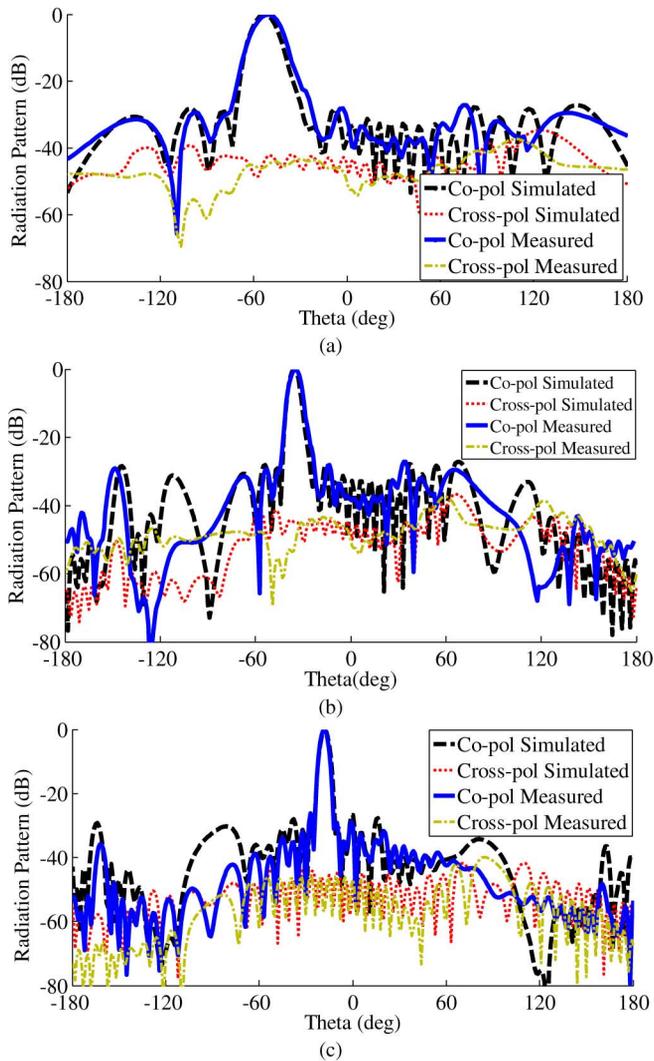

Fig. 7. Measured and simulated co- and cross polarization of the radiation patterns for the proposed antenna at (a) 9, (b) 10, and (c) 11 GHz.

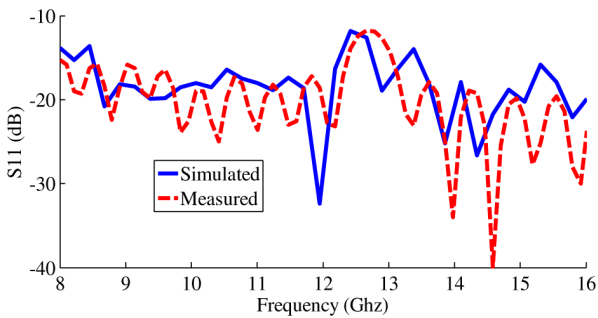

Fig. 8. Measured and simulated $S_{11}$ parameter for the LWA based on sinusoidal RSIW proposed in Table I.

an input reflection coefficient level lower than 10 dB is measured. The measurement is in a good agreement with HFSS simulations.

## VI. CONCLUSION

The two-layer SIW long slot leaky-wave antenna is studied and fabricated with a controllable SLL and low cross polarization. Using a sinusoidal ridge makes radiation possible for a slot located in the center of the upper plane of the SIW and, at the same time, reduces the cross polarization. On the other hand, the phase and leakage constants can be controlled by varying the ridge's parameters. As a result, a continuous angular scanning of $-90°$ to $40°$, over the wide bandwidth of 8.5–20 GHz is observed. The proposed structure is more complicated and more expensive in comparison to similar conventional structures, but possesses a low cross polarization. In addition, the phase constant is kept fixed when the leakage rate is varied. Measurement results also confirm a good SLL and low cross polarization as predicted in HFSS simulations, as well as the beam steering in frequency.